\documentclass[letterpaper, 10 pt, conference]{ieeeconf}  

\IEEEoverridecommandlockouts                              

\usepackage{graphics} 
\usepackage{epsfig} 
\usepackage{amsmath} 
\newcommand{\RNum}[1]{\uppercase\expandafter{\romannumeral #1\relax}}
\usepackage{multirow}
\usepackage[noadjust]{cite}
\title{\LARGE \bf
 Uncertainty-weighted Multi-tasking for $T_{1\rho}$ and $T_2$ Mapping in the Liver with Self-supervised Learning
}

\author{Chaoxing Huang$^{1*}$ Yurui Qian$^{1}$ Jian Hou$^{1}$ Baiyan Jiang$^{1,2}$ Queenie Chan$^{3} $ \\ Vincent WS Wong$^{4}$  Winnie CW Chu$^{1}$  Weitian Chen$^{1}$
\thanks{*Corresponding email: chaoxing.huang@link.cuhk.edu.hk}
\thanks{$^{1}$Department of Imaging and Interventional Radiology, The Chinese University of Hong Kong, Hong Kong SAR, China}%
\thanks{$^{2}$Illuminatio Medical Technology Ltd, Hong Kong SAR, China}%
\thanks{$^{3}$Philips Healthcare, Hong Kong SAR, China}%
\thanks{$^{4}$Department of Medicine \& Therapeutics, The Chinese University of Hong Kong, Hong Kong SAR, China}%
}

\begin{document}

\maketitle
\thispagestyle{empty}
\pagestyle{empty}

\begin{abstract}
Multi-parametric mapping of MRI relaxations in liver has the potential of revealing pathological information of the liver. A self-supervised learning based multi-parametric mapping method is proposed to map $T_{1\rho}$ and $T_2$ simultaneously, by utilising the relaxation constraint in the learning process. Data noise of different mapping tasks is utilised to make the model uncertainty-aware, which adaptively weight different mapping tasks during learning. The method was examined on a dataset of 51 patients with non-alcoholic fatter liver disease. Results showed that the proposed method can produce comparable parametric maps to the traditional multi-contrast pixel wise fitting method, with a reduced number of images and less computation time. The uncertainty weighting also improves the model performance. It has the potential of accelerating MRI quantitative imaging.

\indent \textit{Clinical relevance}— This study establishes a potential way for accelerating multi-parametric mapping in quantitative magnetic resonance imaging and facilitate their clinical applications. 
\end{abstract}

\section{INTRODUCTION}

\par $T_{1\rho}$ and $T_2$ are two important biomarkers in quantitative MRI (qMRI) for liver pathological studies\cite{serai2021basics,takayama2022diagnostic}. In  multi-tasking (multi-parametric mapping) scenarios, the simultaneous mapping of $T_{1\rho}$and $T_2$  acquires multiple $T_{1\rho}$-weighted images and $T_2$-weighted images within a single breath-hold and the parametric maps are fitted separately\cite{chen2017simultaneous,arihara2022evaluation,thomaides2020multiparametric}. It is desirable to quantify different parametric maps at the same time with a reduced number of MR contrast since it can reduce the scan time and potentially improve the quantification accuracy. 
\par Deep learning has been used as an advanced mapping technique in quantitative MRI to map a reduced number of contrast images or undersampled k-space data to the parametric maps\cite{feng2022rapid,liu2022highly}. While most of the previous works focus on single parametric mapping from only one kind of MR contrast, learning-based multi-parametric mapping has  gained interests recently. Qiu et al.\cite{qiu2022multiparametric} proposed a fully supervised deep learning framework to infer $T_{1}$ and $T_2$ maps of brain simultaneously  from $T_1$ and $T_2$ contrasts. Similarly, Saez et al.\cite{moya2021deep} trained a network from synthetic data in a supervised way and map the parametric maps of $T_{1}$ and $T_2$ of brain.  Li et al.\cite{li2022supermap} used supervised learning method with relaxation constraint to map $T_{1\rho}$ and $T_2$ of knee from a reduced number of undersampled contrasts. All those methods are in a supervised way, which relies on high quality  labelled data. Previous work on learning-based liver parametric mapping shows that supervised learning does not provide satisfactory results as the label outside liver parenchyma are noisy\cite{huang2022uncertainty}. It is not uncommon for the scan protocol to sacrifice the data quality outside the parenchyma to ensure a reliable relaxation quantification in the liver. On the other hand, those learning-based multiparametirc mapping methods simply treat each mapping task equally while ignoring how different mapping tasks contribute to the whole learning process of the model in different ways. This could be problematic as treating different tasks equally in  multi-task learning can sometimes degrade the performance of a single task compare to its single task learning counterpart\cite{vandenhende2021multi}. The intensity scale of different MR constrasts and data noise of different mapping tasks may varies, making the difficulties of learning each parametric mapping different. This could create a bias between mapping tasks during learning. It is also pointed out by Wang et al.\cite{li2022artificial} that taking the noise  and uncertainty of the MRI data into consideration is an open challenge in AI application in multi-parametric MRI. How to better integrate the learning of different mapping tasks in qMRI multi-tasking by utilising the data noise remains to be explored.
\par To tackle the aforementioned problems, we propose a self-supervised multi-parametric mapping method from a reduced number of MR contrasts, which alleviates the needs of ground-truth data during training. We also leverage the concept of  uncertainty loss weighting from multi-task learning in our learning algorithm, which utilises the data noise to exploit  suitable contribution of different mapping tasks during learning.

\section{Materials and Method}
\subsection{Data Acquisition and Dataset}
\par Our in vivo studies were conducted with the approval of the institute. The scans were conducted on a 3.0 T MRI scanner (Philips Achieva, Philips Healthcare,Best, Netherland). The RF transmitter was a body coil and a 32-channel cardiac coil was the receiver. Our pulse sequence can acquire  $T_{1\rho}$ and $T_2$ weighted images for $T_{1\rho}$ and $T_2$ mapping  within a single breath-hold\cite{chen2017simultaneous}. Pencil-beam volume shimming box was placed on the right lobe of the liver to reduce the $B_0$ field inhomogeneity. The $B_1$ field inhomogeneity was reduced using dual
transmit and vendor-provided RF shimming. $T_{1\rho}$-weighted images were acquired under the time of spin-lock (TSL) of 0, 10, 30, and 50 ms; $T_2$-weighted images were acquired under the $T_2$ preparation time (TP) of 0, 20, 40, and 60 ms; The $T_2$ preparation time used for $T_2$ fitting was corrected by subtracting the total refocusing time and was 0, 18.2, 34.6, and 51.0 ms, respectively. The $T_{1\rho}$ -weighted image acquired with TSL = 0 and the $T_2$-weighted image acquired with TP = 0 shared the same image (referred as “shared image” in the following context)\cite{li2014simultaneous}. The protocol acquired three slices of data from each subject. The scan time to collect data for each slice was around 16 s. Detailed imaging parameters configuration are in Table \RNum{1}. 
\begin{table}[h]
\caption{Imaging Parameters Configuration}
   \begin{center}
\begin{tabular}{||c c ||} 
 \hline
 Parameters & Setting \\ [0.5ex] 
 \hline\hline
 Resolution & $1.5\times 1.5$ $ mm^2$  \\ 
 \hline
 Slice Thickness & 7 mm  \\
 \hline
 Time of Repetition & 2000 ms  \\
 \hline
 Frequency of Spin-lock & 400 Hz \\ [1ex] 
 \hline
\end{tabular}
\end{center}
\end{table}
\par The retrospective data of 51 patients with non-alcoholic fatty liver disease was used as the dataset. We followed a three-fold cross validation scheme, with the data of 17 patients in each fold.

\subsection{Method}
We first model the learning-based multi-parametric mapping from a probabilistic perspective and derive the loss function in a supervised way. Then we adopt it into the self-supervised form. 
\subsubsection{Multi-parametric mapping likelihood}
\par Let us denote the output of  the multi-parametric mapping neural network as $\mathbf{f^W(x)}$, with weights $\mathbf{W}$ on input MR contrasts $\mathbf{x}$.  The multi-parametric mapping likelihood can then be defined as $p(\mathbf{T_{1\rho}},\mathbf{T_2}|\mathbf{f^W(x)})$, where $\mathbf{T_{1\rho}}$ and $\mathbf{T_2}$ are the ground-truth parametric maps.
\par We further factorise our output following \cite{kendall2018multi}, and have the likelihood of the multi-parametric mapping in the following form:
\begin{equation}
    p(\mathbf{T_{1\rho}},\mathbf{T_2}|\mathbf{f^W(x)}) = p(\mathbf{T_{1\rho}}|\mathbf{f^W(x)})p(\mathbf{T_2}|\mathbf{f^W(x)})
\end{equation}
We assume the distribution of each factorized likelihood as a Laplacian distribution, and minimise the following negative log likelihood of the multi-parametric mapping:
\begin{equation}
 \begin{split}
   -\log p(\mathbf{T_{1\rho}},\mathbf{T_2}|\mathbf{f^W(x)}) & \propto \frac{|\mathbf{T_{1\rho}-\mathbf{f^W(x)}}|}{\mathbf{\sigma_1}} + \frac{|\mathbf{T_{2}-\mathbf{f^W(x)}}|}{\mathbf{\sigma_2}} \\
  & + \log(2\mathbf{\sigma_1}) + \log(2\mathbf{\sigma_2})
 \end{split}
\end{equation}
where $\mathbf{\sigma_1}$ and $\mathbf{\sigma_2}$ are the scale parameters of different parametric map, respectively. The scale parameter is equivalent to the standard deviation of a Gaussian distribution. Eq(2) is our initial derived objective function to be minimised. Note that the uncertainty terms (scale parameters) are learnable terms, which enable an adaptive uncertainty weighted loss during training. More specifically, this utilises the data noise to automatically tune the contribution of different mapping tasks in the learning process. If the data noise of one of the relaxation mappings is large, the L1 norm at the numerator will be large as the model has more difficulties in learning a good mapping. Consequentially, the uncertainty at the denominator becomes larger to suppress the loss value, which guides the model to put less importance on those noisy data during learning. This adaptive weighting provides more flexibility than manual hard weighting in integrating the information of different measurement in training. 
\par The data uncertainty can be further divided into two categories, the heteroscedastic uncertainty (HETEU) and the homoscedastic uncertainty (HOMOU)\cite{kendall2018multi,bragman2018uncertainty}. The former is dependent on a specific input, and it is usually modeled as an additional output tensor with the same dimension as the output variable in deep learning. The latter is independent of a specific input, while it is task-dependent as it captures the general data uncertainty of the training data of a certain mapping task. It is modelled as a learnable constant during training. We study both case in this work. 
\subsubsection{Leveraging self-supervised learning }
We first briefly introduce the relaxation constraints in the mono-exponential decay model of $T_{1\rho}$ and $T_2$ imaging:
\begin{equation}
    \mathbf{I}(TSL_i) = \mathbf{I}(TSL_j)\exp(\frac{TSL_j-TSL_i}{\mathbf{T_{1\rho}}})
\end{equation}
\begin{equation}
    \mathbf{M}(TP_m) = \mathbf{M}(TP_n)\exp(\frac{TP_n-TP_m}{\mathbf{T_{2}}})
\end{equation}
where $\mathbf{I}$ and $\mathbf{M}$ stands for $T_{1\rho}$-weighted image and $T_2$ weighted image respectively, and $i,j,m,n$ are the index of different dynamic scans of the same slice.\\
Since ground-truth maps are not available in the self-supervised learning settings, we replace the L1 norm in the numerator of the original derived loss function as a signal reconstruction term that complies the above relaxation constraints, and the objective function can further be written as:
\begin{equation}
 \begin{split}
   L & = \frac{|\mathbf{I}(TSL_i) - \mathbf{I}(TSL_j)\exp(\frac{TSL_j-TSL_i}{\mathbf{\widehat{ T_{1\rho}}}})|}{\mathbf{\sigma_1}} \\
   & + \frac{|\mathbf{M}(TP_m) - \mathbf{M}(TP_n)\exp(\frac{TP_n-TP_m}{\widehat{\mathbf{T_{2}}}})|}{\mathbf{\sigma_2}} +  \log(2\mathbf{\sigma_1}) \\ 
   & + \log(2\mathbf{\sigma_2})
 \end{split}
\end{equation}
where $\mathbf{\widehat{ T_{1\rho}}}$ and $\mathbf{\widehat{ T_{2}}}$ are the predicted parametric maps from the neural network.  In practice, all possible pairs of constraint (($i,j$) or ($m,n$)) are constructed and back-propogated to update the network parameters during learning. 
\subsubsection{Network setting}
We adopt a similar U-Net architecture for parametric mapping as in \cite{huang2022uncertainty}, in which the output layer has two channels, one for $T_{1\rho}$ map and one for $T_2$ map. The input is a three-channel tensor consists of the shared image, a $T_{1\rho}$ contrast and a $T_2$ contrast. As for the case of estimating HETEU, an additional decoder branch was added to output the uncertainty. The additional decoder branch has the same architecture as the decoder branch for parametric mapping. The illustration are shown in Fig. 1.
\begin{figure}[h]
    \centering
    \includegraphics[scale=0.4]{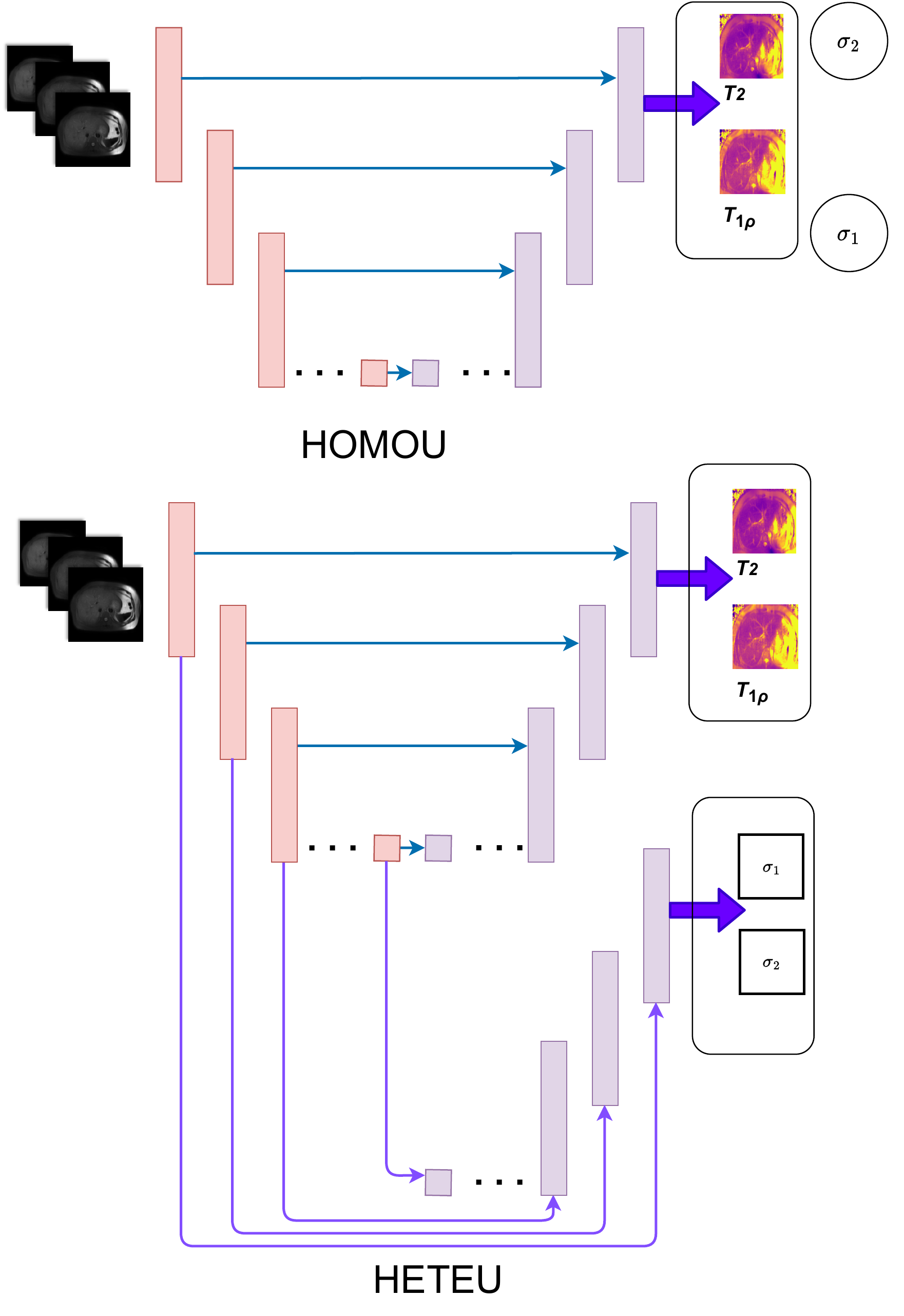}
    \caption{The illustration of the network settings. HOMOU stands for training with homoscedastic uncertainty and HETEU stands for training with Heteroscedastic uncertainty.}
    \label{haha}
\end{figure}
\section{Experiments and Results}
\subsection{Evaluation metric} We evaluate the performance in the ROI as previous works \cite{huang2022uncertainty,li2022supermap}, by computing the pixel-wise mean absolute error between the inference maps and the reference maps in the ROI. We refer it as ROI Mean Absolute Error (RMAE). The ROI is manually drawn on the right lobe of the liver to cover the parenchyma as much as possible while avoiding large vessels and bile-ducts. The drawing was conducted before any fitting to ensure the evaluation fairness. We used parametric maps fitted by four $T_{1\rho}$ contrasts and four $T_2$ contrasts using the non-linear least square fitting method as the reference map. Note that the area outside the parenchyma is not taken into account as its relaxation values from the reference maps are not reliable due to the application of the localized shimming on the right lobe of the liver.
\subsection{Implementation details} 
The experiments were carried out using Python 3.7 and Pytorch 1.10 framework\cite{paszke2019pytorch} , with one Nvidia GTX 1080ti GPU and 40 E5-2630 CPUs. All images were resized as 256 × 256, and data augmentation was applied with random slight rotation and translation. During training and testing, we constructed three combinations of input with images from different TSL or TP. The combinations were as follow:\\
$[I(TSL=0),I(TSL=10 ms),M(TP=18.20 ms)]$\\
$[I(TSL=0),I(TSL=30 ms),M(TP=34.60 ms)]$\\
$[I(TSL=0),I(TSL=50 ms),M(TP=51.00 ms)]$\\
Batch size was 4 and the learning rate was 5e-4. ADAM \cite{kingma2014adam} was used as the optimizer with a weight decay of 1e-4. The two learnable constant in HOMOU were initialized as 1. Each fold of training took around 8 hours for 300 epochs and early stopping was applied. 
\subsection{Comparison study}
We compare our proposed method with the following models:\\
\subsubsection{Two-point} The logarithm of the quotient between the shared image and the corresponding $T_{1\rho}$ weighted image or the $T_2$ weighted image is taken to get the parametric maps in a closed form. 
\subsubsection{Single task with single modality (STSM)} Two separate self-supervised networks that map $T_{1\rho}$ and $T_2$ map was trained respectively.  Each mapping task follows the "Baseline" method in \cite{huang2022uncertainty}. The input consists of the shared image and the corresponding contrast (shared image $+ T_{1\rho}$ contrast for $T_{1\rho}$ mapping or shared image $+ T_2$ contrast for $T_2$ mapping). The loss function is the L1 norm for signal reconstruction in self-supervised learning based on the constraint shown in Eq (3) or Eq (4).  
\subsubsection{Single Task (ST)} This is similar to STSM, except for the input. The input is the three-channel tensor  as that of our proposed method.
\subsubsection{Supervised Learning (SL)} The multi-parametric mapping network is trained in a supervised way similar to those previous work\cite{qiu2022multiparametric,moya2021deep}. The input of the network is the same as that in our proposed method and the ground-truth for supervision were the reference maps fitted by four images. The loss function is the sum of the L1 norm of both mapping tasks.
\subsubsection{Baseline} The network is trained in a self-supervised way without those uncertainty terms in the loss. 
\par The results are shown in Table \RNum{2}. It can be seen that the self-supervised baseline model outperforms the traditional two point fitting, supervised learning model and those self-supervised single task models. By adding HETEU or HOMOU during  training, the performance of the model can be further improved to the level of around 3.30 ms. The performance of HETEU and HOMOU are close to each other, and details will be provided in the discussion section. 
\par Fig. 2 shows examples of the fitted $T_{1\rho}$ maps and $T_2$ maps. As is shown, the Two-Point method produced very noisy results and the SL method results are poor at revealing the anatomical information with oversmoothing effect. This is aligned with those result in supervised single parmetric mapping reported in\cite{huang2022uncertainty}. The maps produced by our proposed method demonstrate a general good agreement with the reference maps in the right lobe of the liver parenchyma in the ROI.

\begin{figure*}[htp]
    \centering
    \includegraphics[width = 18cm]{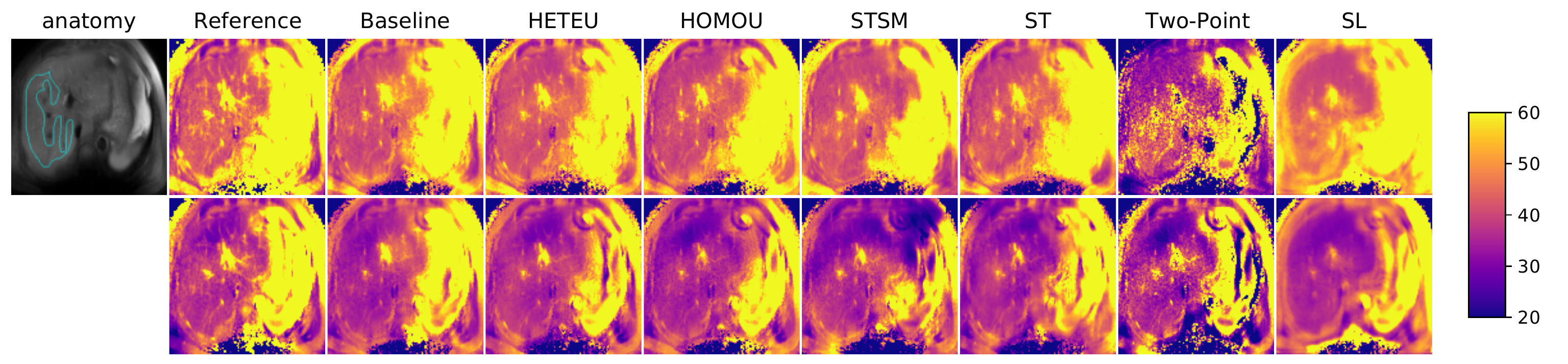}
    \caption{A typical example of the predicted maps. The green countour is the region of intersts. The maps in the first row are the $T_{1\rho}$ maps and the maps in the second row are the $T_2$ maps. The unit of the color bar is in ms. }
    \label{hah}
\end{figure*}

\begin{table}[h]
\caption{RMAE of different models in the comparison study. The unit is in ms.}
   \begin{center}
\begin{tabular}{||c c c ||} 
 \hline
 Models & $T_{1\rho}$ & $T_2$ \\ [0.5ex] 
 \hline\hline
 Two-point&  7.51& 5.21 \\ 
 \hline
 STSM & 3.54 & 4.08 \\
 \hline
 ST & 3.54  & 3.87 \\
 \hline
 SL &  5.11 & 4.12 \\ 
 \hline
 Baseline & 3.45 & 3.51 \\ 
 \hline
 Baseline + HETEU & 3.32  & \textbf{3.32} \\
 \hline
 Baseline + HOMOU& \textbf{3.28} & 3.33 \\ [1ex] 
 \hline
\end{tabular}
\end{center}
\end{table}
\subsection{Effectiveness of adaptive weighting}
\par Two experiments were conducted in this session. For the first experiment, we compared the results of our proposed method with those baseline models with different manual tuned weights between the $T_{1\rho}$ contrast reconstruction term and the  $T_{2}$ contrast reconstruction term. The loss function follows the form as: $\lambda_a L_a + \lambda_b L_b $, where $L_a$ and $L_b$ stand for the $T_{1\rho}$ contrast and $T_2$ contrast reconstruction loss term respectively, and $\lambda_a+\lambda_b = 1$ . For the second experiment, we investigate if the adaptive weighting can handle those situation with scaling variation and data corruption. Specifically, we multiply the signal scale of $T_2$ contrast in the reconstruction term with a very large factor or apply random motion (rotation and translation) to $T_2$ contrast in the loss function.
\par Fig. 3 shows the results of the first experiment. It can be seen that the performance of the self-supervised network is sensitive to manual weighing, and the mapping task performance bias can be seen in some of the weighting scenarios. On the other hand, the uncertainty weighted adaptive weighting give an overall improved performance and the performance bias is not obvious.  
\par In Table \RNum{3}, the results show that applying signal scale imbalance and motion corruption to $T_2$ contrast in the loss function give inferior $T_{1\rho}$ results for the Baseline model, compared with the baseline $T_{1\rho}$ results in Table \RNum{2}. By applying the uncertainty-weighted loss, the performance degradation can be relieved. Note the performance of $T2$ is not reported in the motion scenario as the $T_2$ data were corrupted. It is also noticeable that the HOMOU method produce a better results than the HETEU method, detailed discussion will be provided. 
\begin{figure}[htp]
    \centering
    \includegraphics[width=8cm]{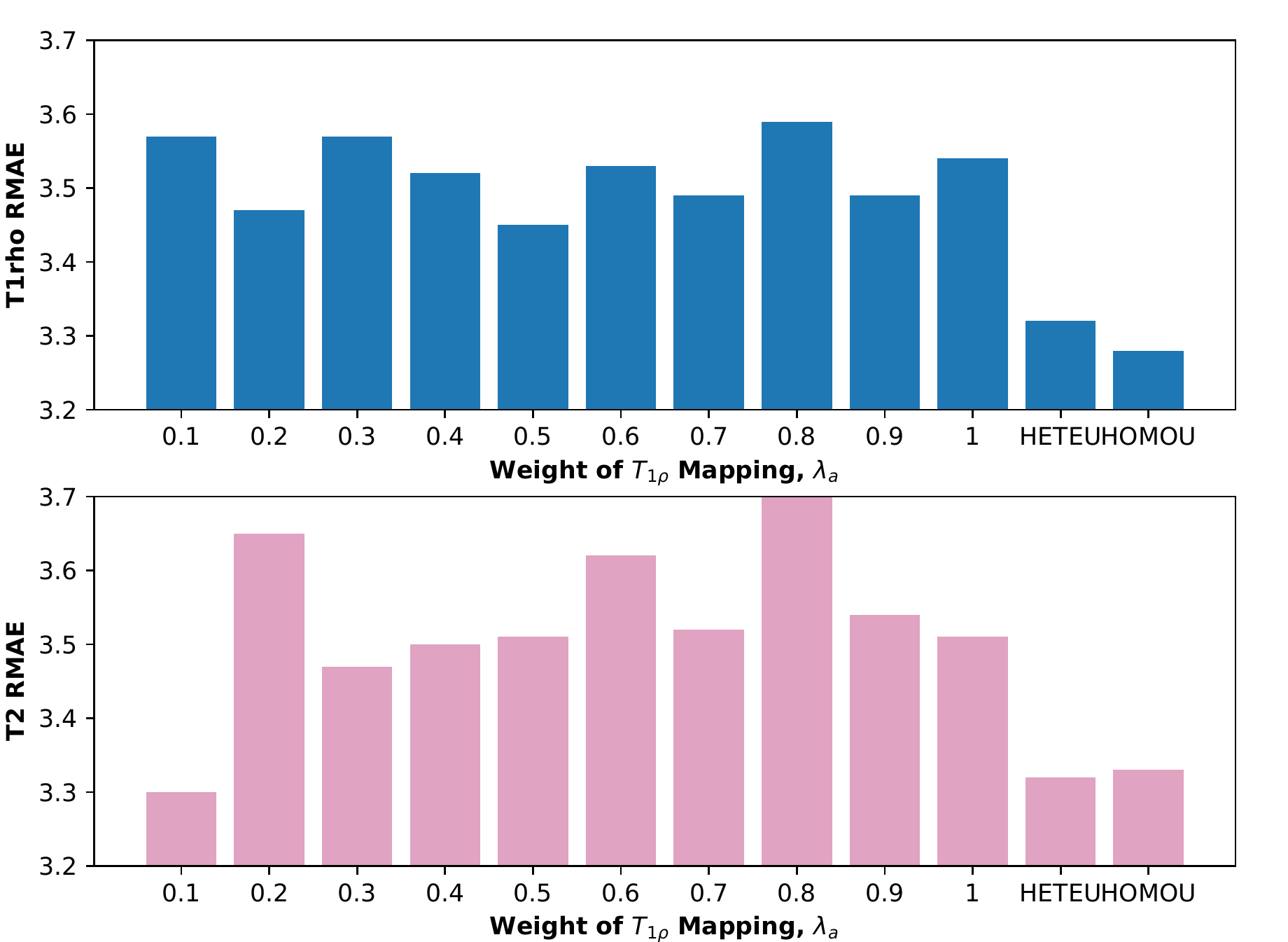}
    \caption{Performance of different loss weighting in self-supervised multi-parametric mapping }
    \label{fig:galaxy}
\end{figure}

\begin{table}[h]
\caption{RMAE of the models under scaling and motion corruption. The unit is in ms.}
   \begin{center}
\begin{tabular}{||c c c ||} 
 \hline
$\mathbf{\times 100}$ & $T_{1\rho}$ & $T_2$ \\ [0.5ex] 
 \hline
 Baseline & 3.62 & 3.44 \\ 
 \hline
 Baseline + HETEU & 3.53  & 3.49 \\
 \hline
 Baseline + HOMOU& 3.45 & 3.44\\ [1ex] 
 \hline
 \hline
  $\mathbf{\times 10000}$ &  $T_{1\rho}$ & $T_2$   \\ [0.5ex] 
  \hline
  Baseline & 3.88 & 3.67 \\ 
 \hline
 Baseline + HETEU & 3.56  & 3.40 \\
 \hline
 Baseline + HOMOU& 3.44 & 3.45 \\ [1ex] 
 \hline
 \hline
  \textbf{Motion} &  $T_{1\rho}$ & $T_2$  \\ [0.5ex] 
  \hline
  Baseline & 3.55 & - \\ 
 \hline
 Baseline + HETEU & 3.44 & - \\
 \hline
 Baseline + HOMOU& 3.39 & -\\ [1ex] 
 \hline
 
\end{tabular}
\end{center}
\end{table}

\subsection{Computation time comparison}
\par We compared the computation time of the proposed model using three images and the least square fitting method using four images for one single forward pass using the same computational resources and programming framework, which is mentioned in the implementation details section. The computation time was obtained by averaging the forward pass computation time over all the predictions in the data-set. Table \RNum{4} illustrates that the proposed method achieved a shorter computation time than that of the traditional pixel wise fitting method. The former can simultaneously produce two parametric maps at one forward pass while the latter can only produce one parametric map at one forward pass.

\begin{table}[h]
\begin{center}
\caption{Average computation time of the pixel-wise least square method and the proposed method}
\begin{tabular}{ |c|c| } 
\hline
Method & Computation Time \\
\hline
Pixel-wise least square fitting & 0.088 s  \\ 
Proposed & 0.017 s \\ 

\hline
\end{tabular}
\end{center}
\end{table}

\

\section{Discussion}
  \par The results demonstrate that our proposed learning-based method can produce comparable multi-parametric mapping results to the standard multi-image pixel-wise fitting method, by using less images and computation time. From a practical point of view, this could potentially improve the efficiency in large scale qMRI study, as the acquisition time and the post-processing time are both reduced.
  \par Our studies also demonstrated the benefits of the the uncertainty-based adaptive weighting. It improves the model performance by utilising the data noise of different tasks in the multi-parametric mapping to automatically exploit a better contribution mechanism from different mapping tasks. This is beneficial as it saves the time of manual weight tuning. Future multi-sequences or multi-sites study may also see benefits, as the data noise variation  problem can be more significant in those scenarios. 
  \par It is also noticeable that the performance of the HETEU and HOMOU method are similar, and the HOMOU achieves a slightly better performance under the case of scaling variation and motion corruption.  The HETEU model  captures the pixel-wise weighting  while the HOMOU model  capture the general weighting of the two task globally. The former may includes redundant spatial information in learning the adaptive  weighting, and the weighting strategies may overfit on the data. The latter learns the task-specific weighting, and it may reflect a general contribution mechanism of different parametric mapping tasks. 
  \par In this work, we factorise the likelihood of the multi-parametric mapping with the assumption that the distribution of $T_{1\rho}$ mapping and the distribution of $T_2$ mapping are independent from each other. While it is common to apply the task independent assumption in multi-task learning, we believe it is worth exploring the correlation of $T_{1\rho}$ and $T_2$ mapping in future learning-based multi-parametric mapping research. Recent work on hepatic iron in the liver has demonstrated their correlation bio-physically\cite{qian2022characterization}. Future research will focus on applying multi-variate multi-task learning\cite{russell2021multivariate} to learn the correlation between different mapping tasks.  

\section{CONCLUSIONS}
Our proposed uncertainty-weighted learning-based multi-parametric mapping method is able to simultaneous map $T_{1\rho}$ and $T_2$ in the liver from a reduced number of contrasts. The uncertainty-weighted learning improves the  performance of the mapping model by utilising the data noise of different mapping tasks. Future work on learning correlation between different mapping tasks in multi-parametric mapping is required.

\addtolength{\textheight}{-12cm}   




\section*{ACKNOWLEDGMENT}

This study was supported by a grant from the Research Grants Council of the Hong Kong SAR (Project GRF 14201721), a grant from the Innovation and Technology Commission of the Hong Kong SAR (Project No.MRP/046/20x).


\bibliographystyle{IEEEtran}
\bibliography{ref.bib}

\end{document}